\documentclass[conference]{IEEEtran}

\usepackage{orcidlink}
\usepackage{hyperref}
\usepackage{xcolor}
\usepackage{lipsum}
\usepackage{multirow}
\usepackage{amsfonts}
\usepackage{mathtools}
\usepackage{verbatimbox}
\usepackage{enumitem}
\usepackage{pifont}
\usepackage{rotating}
\usepackage{booktabs}
\usepackage{pdflscape}
\usepackage{caption}
\usepackage{enumitem}

\usepackage{tikz}
\usetikzlibrary{shapes.geometric, arrows,arrows.meta,positioning,calc}
\newcommand*\circled[1]{\raisebox{.5pt}{\textcircled{\raisebox{-.9pt} {#1}}}}

\usepackage{algorithm}
\usepackage{algpseudocode}

\newcommand{\oracle}[1]{\textbf{#1}}
\newcommand{\nocompile}[1]{}

\begin{document}

\nocompile{
\author{
Daniel Engel\inst{1}\orcidID{0009-0004-0989-3869} \and
Freek Verbeek\inst{1}\orcidID{0000-0002-6625-1123} \and
Pranav Kumar\inst{2}\orcidID{0009-0005-3619-2969} \and
Binoy Ravindran\inst{2}\orcidID{0000-0002-8663-739X}
}

\authorrunning{Engel et al.}

\institute{
Open University, Heerlen, The Netherlands \\
\email{\{daniel.engel, freek.verbeek\}@ou.nl}
\and
Virginia Tech, Blacksburg, USA \\
\email{\{pranavkumar, binoy\}@vt.edu}
}
}

\author{

\IEEEauthorblockN{Daniel Engel, Freek Verbeek}
\IEEEauthorblockA{
Open University, Heerlen, The Netherlands \\
Email: \{daniel.engel, freek.verbeek\}@ou.nl
}

\IEEEauthorblockN{Pranav Kumar, Binoy Ravindran}
\IEEEauthorblockA{
Virginia Tech, Blacksburg, USA \\
Email: \{pranavkumar, binoy\}@vt.edu
}

}

\title{Adding Compilation Metadata To Binaries To Make Disassembly Decidable}

\maketitle

\renewcommand\thesection{\arabic{section}}
\renewcommand\thesubsectiondis{\thesection.\arabic{subsection}}
\renewcommand\thesubsubsectiondis{\thesubsectiondis.\arabic{subsubsection}}

%% =============================================================================
%% == ABSTRACT =================================================================
%% =============================================================================

\begin{abstract}
The binary executable format is the standard method for distributing and
executing software.  Yet, it is also as opaque a representation of software
as can be.  If the binary format were augmented with metadata that provides
security-relevant information, such as which data is intended by the
compiler to be executable instructions, or how memory regions are expected
to be bounded, that would dramatically improve the safety and
maintainability of software. In this paper, we propose a binary format that
is a middle ground between a stripped black-box binary and open source. We
provide a tool that generates metadata capturing the compiler’s intent and
inserts it into the binary. This metadata enables lifting to a correct and
recompilable higher-level representation and makes analysis and
instrumentation more reliable. Our evaluation shows that adding metadata
does not affect runtime behavior or performance. Compared to DWARF, our
metadata is roughly 17\% of its size. We validate correctness by compiling
a comprehensive set of real-world C and C++ binaries and demonstrating that
they can be lifted, instrumented, and recompiled without altering their
behavior.
  \end{abstract}
  
\nocompile{


}

%% =============================================================================
%% == INTRODUCTION =============================================================
%% =============================================================================

\section{Introduction}\label{sec:introduction}

%% ========== Context (binaries are unanalyzable)
%
The binary format (e.g., ELF but the same applies to MACHO or PE) is the
de-facto standard method of distributing software. It also is, however,
essentially a black-box. Without source code available, it is hard to analyze a
binary, to apply patches to it, to test or to fuzz it. This is the cause of
various security issues, undetected or unpatched vulnerabilities, and exploits.
If we would distribute software in a binary format that is analyzable,
editable, instrumentable, and fuzzable, that could dramatically impact the
safety of our software. We thus argue software should not be distributed as
black-boxes. However, distributing source code is not always an option either:
third-party closed source software is ubiquitous. Enterprises need the
capability of distributing software without publishing their source code or
revealing proprietary internals. The question then is: what would a
\emph{middle ground binary format} look like that on one hand is ``easier'' to
analyze and test, and ``more instrumentable'', but that on the other hand does
not require one to distribute the source, nor makes it substantially easier to
retrieve the source through decompilation?

%% ========== Answering the question: what is needed to solve this?
%
First, let us consider this question practically with some examples. In order
to do an analysis (e.g., verify memory-related properties or the absence of
overflows), one needs to have bounds on the sizes of memory regions, and one
needs to be able to traverse the binary, which requires control flow graphs
(CFG) of each
function~\cite{balakrishnan2004analyzing,song2008bitblaze,wang2009intscope,brumley2011bap,djoudi2015binsec}.
In order to apply a patch, or to instrument a binary for fuzzing, one needs to
be able to insert instructions and data into a
binary~\cite{ben2020efficient,dinesh2020retrowrite}. Currently, a stripped
Executable and Linkable Format (ELF) file does not enable any of these actions.
However, for these kinds of actions, one does \emph{not} necessarily need
variable names, type information, or anything related to the structure of the
original source code.

To summarize these requirements, we argue the need for a middle ground binary
format that is 1.) \emph{traversable}, 2.) \emph{memory-structured}, and 3.)
\emph{instrumentable}. Traversability means that given the current instruction,
we have a reasonably tight overapproximation of what the next instruction to be
executed is. This is a challenge, since in binaries the address of the next
instruction may be computed dynamically at run-time (indirections).
Memory-structured means that compound contiguous memory segments such as stack
frames (where local data is stored) and data sections (where global data is
stored) can be decomposed into bounded independent memory regions. Without this
property, analysis tools must treat memory segments as single undifferentiated
arrays of bytes, thwarting verification of memory-related properties.
Instrumentable means that the binary can be edited by adding instructions and
data.

Based on these requirements,  we argue that a middle ground binary format at
least should provide \emph{symbolized
assembly}~\cite{wang2017ramblr,verbeek2024verifiably}. Symbolization replaces
absolute addresses and offsets with symbolic labels. If the text- and data
sections of a binary are symbolized, this enables insertion of new instructions
and data, required for patching and instrumentation. Symbolization, however, is
not enough. Symbolized assembly still can be untraversable due to indirections,
and need not be memory-structured: information on which parts of a memory
segment are to be considered independent can be lost. We will define
\emph{fully symbolized assembly} as symbolized assembly with additional
characteristics that ensure the above three properties.

Retrieving fully symbolized assembly from a normal binary requires the solving
of various problems that are \emph{undecidable}. First of all, it requires
disassembly, which in itself is already undecidable. Subsequently, it requires
at least indirection resolving, function boundary detection, call graph
construction, per-function termination analysis, and CFG construction just to
obtain a traversable intermediate representation (IR). Then, recovering a
structured memory model for stack frames and data sections requires advanced
forms of pointer and shape analysis, or analysis based on probabilistics or
heuristics~\cite{balakrishnan2004analyzing,balakrishnan2005recovery,kinder2010precise,navas2012signedness,zhang2021osprey,rose2024modeling}.
Simply establishing which calling convention was used during compilation of the
binary already requires a data flow analysis. All of these problems are
undecidable~\cite{rice1953classes}. However, all of the answers were known at
build time.

To emphasize this point more, we state that solving the above problems is not
just theoretically undecidable, but that in practice state-of-the-art tools
also often
fail~\cite{andriesse2016depth,djoudi2016recovering,pang2021-sok,zhibo20}. This
holds true even though a large body of both academic literature as well as
industrial tooling exists, in the fields of binary analysis, reverse
engineering, and decompilation. Even today's state-of-the-art tools will not
reliably disassemble and lift all binaries, are not capable of resolving all
indirections, and do not produce recompilable code in a trustworthy fashion.
This statement holds even when we only consider benign binaries produced by
compilers. The core of the issue is that these tools have the impossible job of
recovering information that was known -- but lost -- while building the binary.

%% ========== Contribution
%
The contribution of this paper is a binary format we call the \emph{Executable,
Linkable, and Liftable Format (ELLF)}. An ELLF is an ELF, augmented with a
minimum amount of metadata that prevents the need to solve any of the above
problems, or in other words, that makes the above problems decidable. We first
discuss what exact metadata needs to be added to achieve this. We show how this
metadata can be gathered at build time and how it can be added to an ELF
without significant overhead in terms of compilation time and binary file size.
We are agnostic of the compiler settings: one can still compile the code as one
likes (e.g., at any optimization level). We then show that the above problems
are indeed decidable by providing executable and efficient algorithms.
%Finally, we show that various use cases are enabled by distributing ELLFs
%rather then ELFs.\todo{are we doing this?}
To the best of our knowledge, this is the first effort on generating ground
truth information at build time, packing it into a binary format, evaluating
its correctness, \emph{and} demonstrating that it is sufficient for lifting to
a recompilable higher-level IR.

%% ========== State-Of-The-Art (DWARF,ELF)
%
Currently, there are three ways to distribute an ELF: stripped, unstripped, or
augmented with DWARF debugging information. First, note that all of these
formats contain \emph{too little information} to make any of the above problems
decidable. For example, even for unstripped ELF files with DWARF information,
recovering control flow is undecidable. Secondly, note that unstripped ELFs --
with or without DWARF debugging information -- also contain \emph{too much}
information. They contain -- among others -- names of global variables and
internal functions, which is why many vendors choose to distribute stripped
binaries instead. DWARF in particular includes a large amount of information
that is irrelevant in this context and can be considered undesirable in the
final binary, such as mappings from binary locations to their original source
code locations. There is no format that is as close as possible to a stripped
ELF, but that \emph{does} provide crucial and security-relevant information on,
e.g, the intended jump targets of indirections, or on which addresses in the
binary are intended to be executable instructions and which not.

\textbf{Example.} Figure~\ref{fig:disassembly_comparison} shows a comparison of
disassembling a stripped ELF, disassembling an ELF with DWARF, and
disassembling an ELLF.
When only the raw bytes of the binary are available, it is impossible to
distinguish between instructions and raw data bytes, possibly causing
disassemblers to misinterpret them.
With DWARF debugging information, some of these problems can be alleviated as
the entries of functions are often known and values can be compared to known
pointer addresses.
However, this approach is still prone to misinterpretation as soon as text and
data are mixed.
Additionally, if raw values coincide with known addresses, these raw values can
be misinterpreted to be pointers.
In the ELLF, such misinterpretation do not occur, as it contains the exact
instruction locations and marks values as either raw values or pointers.

\begin{figure}
  \centering
  \input{figures/disassembly_comparison.tex}
  \caption{Disassembly comparison.}
    % : stripped binary (top) has no names
    % and misinterprets code, ELF+DWARF (middle) recovers some names and
    % data labels but still misinterprets code, and ELLF (bottom)
    % enables fully symbolized and accurate disassembly.
  \label{fig:disassembly_comparison}
\end{figure}

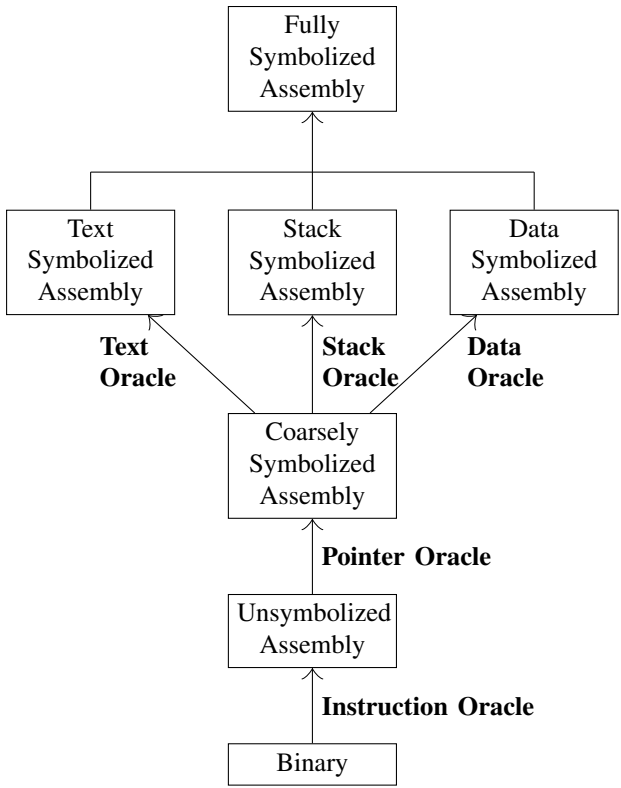
\begin{figure}
  \centering
    \begin{tikzpicture}[]

    \tikzstyle{stage} = [rectangle, text centered, text width=2cm, draw=black]
    \tikzstyle{arrow} = [draw, -{Classical TikZ Rightarrow[length=1.5mm]}]

    \node (binary)   [stage] {Binary};
    \node (assembly) [stage, above=1.0cm of binary]                   {Unsymbolized Assembly};
    \node (coarse)   [stage, above=1.0cm of assembly]                 {Coarsely Symbolized Assembly};
    \node (stack)    [stage, above=1.3cm of coarse]                   {Stack Symbolized Assembly};
    \node (text)     [stage, left=0.7cm of stack] {Text Symbolized Assembly};
    \node (data)     [stage, right=0.7cm of stack] {Data Symbolized Assembly};
    \node (fully)    [stage, above=1.3cm of stack]   {Fully Symbolized Assembly};

    %\node[anchor=north, xshift=1.75cm] at (binary.south) {+ Instruction Oracle};
    %\node[anchor=north, xshift=1.75cm] at (assembly.south) {+ Points-To Oracle};
    %\node[anchor=east] at (coarse.west) {Basic Block Oracle +};
    %\node[anchor=west] at (coarse.east) {+ Variable Oracle};

    \draw [arrow] (binary)   -- node[right] {\oracle{Instruction Oracle}}  (assembly);
    \draw [arrow] (assembly) -- node[right] {\oracle{Pointer Oracle}} (coarse);
    \draw [arrow] (coarse)   -- node[left]  {\begin{tabular}{l}\oracle{Text}\\\oracle{Oracle}\end{tabular}} (text);
    \draw [arrow] (coarse)   -- node[right,xshift=-0.2cm] {\begin{tabular}{l}\oracle{Stack}\\\oracle{Oracle}\end{tabular}} (stack);
    \draw [arrow] (coarse)   -- node[right,xshift=0.25cm] {\begin{tabular}{l}\oracle{Data}\\\oracle{Oracle}\end{tabular}}(data);

    \coordinate (A) at ($(text.north)+(0,0.5cm)$);
    \coordinate (B) at ($(stack.north)+(0,0.5cm)$);
    \coordinate (C) at ($(data.north)+(0,0.5cm)$);
    
    \draw [] (text.north) -- (A);
    \draw [] (stack.north) -- (B);
    \draw [] (data.north) -- (C);
    \draw [] (A) -- (C);
    \draw [arrow] (B) -- (fully.south);

  \end{tikzpicture}
  \caption{Overview of the lifting pipeline. }
  \label{fig:overview-disassembling-pipeline}
\end{figure}

% \begin{figure}[t]
%   \centering
%   \begin{minipage}[t]{0.40\textwidth}
%     \vspace{0pt}
%     \centering
%     \input{figures/disassembly_comparison.tex}
%     \vspace{-0.5cm}
%     \captionof{figure}{Disassembly comparison.}
%     % : stripped binary (top) has no names
%     % and misinterprets code, ELF+DWARF (middle) recovers some names and
%     % data labels but still misinterprets code, and ELLF (bottom)
%     % enables fully symbolized and accurate disassembly.
%     \label{fig:disassembly_comparison}
%   \end{minipage}
%   \hfill
%   \begin{minipage}[t]{0.58\textwidth}
%     \vspace{0pt}
%     \centering
%     \input{figures/pipeline_overview.tex}
%     \vspace{-0.25cm}
%     \captionof{figure}{Overview of the lifting pipeline. }
%     \label{fig:overview-disassembling-pipeline}
%   \end{minipage}
% \end{figure}

%% ========== Contribution summary
%
\noindent
To summarize, the contributions of this work are:
\begin{itemize}[topsep=0pt]
  \item A study of which metadata is needed in order to make a binary traversable, memory-structured, and instrumentable.
  \item A tool to collect said metadata at build time, pack it into an efficient representation, and insert it as a section into the binary, producing ELLF files.
  \item A lifter that consumes an ELLF file and lifts it to correct fully symbolized assembly code.
\end{itemize}

%% ========== Structure and Evaluation
%
%\todo{revisit when paper is finished}
Section~\ref{sec:overview} introduces the structure of the metadata we embed
into binaries and explains how it enables precise disassembling and
symbolization.
Section~\ref{sec:implementation} describes the metadata generation in more
detail, including how the relevant information is extracted from the compiler
outputs, how it is encoded, and how it is later consumed to disassemble
stripped binaries.
Section~\ref{sec:evaluation} presents our evaluation, in which we compile the
LLVM Test Suite and embed ELLF metadata into the binaries.
%Section~\ref{sec:evalutation} presents our evaluation, in which we use a
%modified LLVM toolchain to compile a diverse set of real-world binaries ---
%including GNU Coreutils, binutils and bigger binaries --- with embedded
%metadata.
%
After stripping the binaries, we disassemble them using only our metadata and
the binary itself.
We then reassemble the disassembled output using a standard \texttt{gcc} and
verify that the recompiled binaries match the behavior of the original binaries
across the available test suites.
This validates not just the completeness of the metadata but also the
correctness of the disassembly.
Section~\ref{sec:related-work} discusses related work in extracting
disassembling ground truth from the compiler and in lifting binaries to a
rewritable format.
This paper concludes with a discussion in Section~\ref{sec:conclusion}.
%
%Appendix~\ref{sec:open-science} describes how to access the submission
%artifacts during the double-blind review.

%% =============================================================================
%% == OVERVIEW =================================================================
%% =============================================================================

\section{ELLF Metadata}\label{sec:overview}

Our metadata design is informed by exploring the needs of a lifter capable of
taking a binary and producing fully symbolized assembly. The lifter works in
five steps, each raising the level of abstraction (see
Figure~\ref{fig:overview-disassembling-pipeline}). Each step is in itself
undecidable, and therefore requires an \emph{oracle}: information that makes
the step decidable. Per step, we provide details on what the input and output
is, what oracle is needed and why the oracle is necessary to make the step
decidable. In this section, we provide an \emph{abstract view} of the
information provided by the oracles, or in other words, a specification. The
concrete implementation will differ, mostly because storing the information
naively will take up a lot of space. Section~\ref{sec:implementation} provides
more information on the actual concrete metadata inserted into the binary.

The first two steps have to be applied sequentially. The steps on text-, data-
and stack symbolization can be executed independently in any order. Once all
three have been achieved, this produces fully symbolized assembly.

We use $\mathbb{W}_{64}$ to denote the set of 64-bit words.
Notation $X?$ is used to denote the set $X \cup \bot$ with $\bot$ a fresh element representing ``None''.
Notation $f :: X \rightharpoonup Y$ denotes that $f$ is a partial function from set $X$ to set $Y$.

\begin{figure}[t]
  \centering
  \scalebox{0.75}{%
  \begin{tikzpicture}[]

    \tikzstyle{stage} = [rectangle, text centered, draw=black]
    \tikzstyle{arrow} = [draw, -{Classical TikZ Rightarrow[length=1.5mm]}]

    \node (binary)   [stage] {\boxBinary};
    \node (assembly) [stage, above=1cm of binary]   {\boxUnsymbolized};
    \node (coarse)   [stage, above=1cm of assembly] {\boxCoarseSymbolized};
    \node (text)     [stage, above=1cm of coarse]   {\boxTextSymbolized};

    \node (instr-oracle) [below=0.3cm of binary, xshift=-2.2cm, anchor=north]        {\begin{tabular}{l}\circled{1}~\oracle{Instruction Oracle}\\$\{\mathtt{0x4000, 0x4001, \ldots}\}$\end{tabular}} ;
    \node (ptr-oracle)   [right=0.3cm of instr-oracle.north east, anchor=north west]                               {\begin{tabular}{l}\circled{2}~\oracle{Pointer Oracle}\\$\mathtt{0x4004 \mapsto [\bot,0x4024]}$\\$\mathtt{0x4024 \mapsto 0x4014-0x4024}$\\$\mathtt{0x402C \mapsto 0x401C-0x4024}$\end{tabular}} ;
    \node (text-oracle)  [below=0.5cm of instr-oracle.south west, anchor=north west] {\begin{tabular}{l}\circled{3}~\oracle{Text Oracle}\\$\mathtt{0x4000 \mapsto \mathsf{FunctionStart}}$\\$\mathtt{0x4014 \mapsto \mathsf{BB}}$\\$\mathtt{0x4023\mapsto \mathsf{FunctionEnd}}$\end{tabular}} ;
    \node (data-oracle)  [below=0.1cm of ptr-oracle.south west, anchor=north west]  {\begin{tabular}{l}\circled{4}~\oracle{Data Oracle}\\$\mathtt{0x4004 \mapsto 8}$\\$\mathtt{0x4012\mapsto 8}$\end{tabular}} ;

    \draw [arrow] (binary)     -- node[right] {\circled{1}}  (assembly);
    \draw [arrow] (assembly)   -- node[right] {\circled{2}}  (coarse);
    \draw [arrow] (coarse)     -- node[right] {\circled{3}, \circled{4}}  (text);
  \end{tikzpicture}
  }
  \caption{Example of lifting, with examples of the data provided by the oracles.}
  \label{fig:steps_example}
\end{figure}
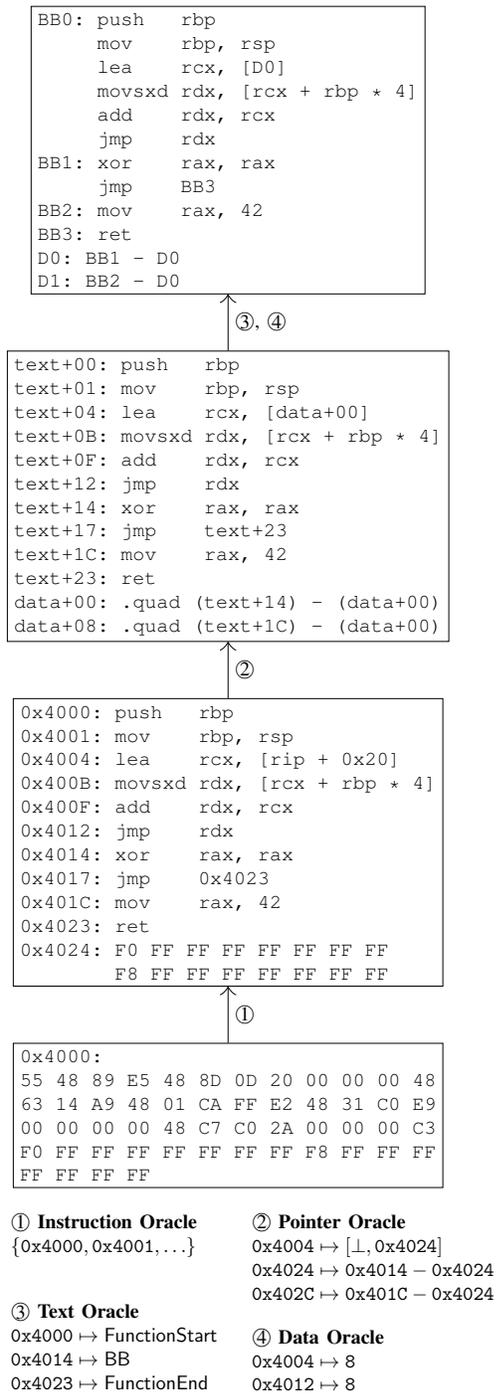

\textbf{Step I: unsymbolized assembly}. Step I takes as input raw data and
outputs disassembled instructions (see Figure~\ref{fig:steps_example}). It is
one of the primary challenges in disassembling to determine the exact locations
of instructions, and this is well-known to be
undecidable~\cite{schwarz2002disassembly,wartell2011differentiating,smithson2013static,wartell2014shingled}.
Although the function boundary information available in DWARF can provide some
guidance, identifying all instruction addresses remains undecidable.

The instruction oracle, abstractly seen, provides a set of addresses that are intended by the compiler to be the starting addresses of executable instructions.
We model the instruction oracle as follows:
\[
	\mathbf{InstructionOracle} :: \{ \mathbb{W}_{64} \}
\]
In other words, the oracle is a set of 64-bit values.

With the instruction oracle in place, lifting to unsymbolized disassembly
becomes straightforward: a disassembler can traverse all addresses provided by
the oracle, disassemble the instruction at the given address or emit the raw
data if no instruction is at that address.

\textbf{Step II: coarsely symbolized assembly}.
Step~II takes as input unsymbolized assembly and lifts it to a representation where all pointers to text- and data-sections are replaced with labels (see Figure~\ref{fig:steps_example}).
The pointer oracle is modeled as:
\[
%         \mathsf{Instr}~[\mathbb{W}_{64}?]
%    \mid \mathsf{Data}~\mathbb{W}_{64}
%    \mid \mathbb{W}_{64} - \mathbb{W}_{64}
  \begin{array}[t]{rl}
    \mathbf{PointerOracle} :: \mathbb{W}_{64} \rightharpoonup
         & \mathsf{Instr}~[\mathbb{W}_{64}?] \\
    \mid & \mathsf{Data}~\mathbb{W}_{64}\\
    \mid & \mathbb{W}_{64} - \mathbb{W}_{64}\\
  \end{array}\\
\]

The pointer oracle is a partial mapping from 64-bit addresses to pointer
information. If the given address $a$ belongs to an instruction, it provides an
ordered list that gives for each operand either $\bot$ (if the operand is not a
pointer), or a pointer value. In Figure~\ref{fig:steps_example}, the
\texttt{lea} instruction has two operands, where the second one is a pointer.
It may also be the case that the given address $a$ falls in some data section,
and that at the location some pointer value $p$ is stored. In that case
$\mathbf{PointerOracle}(a)$ will return $\mathsf{Data}~p$. Finally, it may be
the case that address~$a$ falls in some data section and at that location the
difference between two pointers is stored. This typically happens when
compilers implement jump tables.

Retrieving this oracle from a binary is again undecidable. If the binary is not
RIP-relative, it is undecidable to determine whether, e.g., the value occurring
in the instruction \texttt{mov rax, 0x4000} is a pointer or not. In case the
binary is RIP-relative, this problem does not occur. However, even then the
cases where the origin of a value is the difference between two pointers are
undecidable.

With this oracle in place, symbolization of pointers becomes straightforward. A
disassembler can traverse all instruction operands and replace them by a
symbolic name if they are pointers, and traverse the data sections in similar
fashion as well.

This lifting step produces coarsely symbolized assembly: one can move entire
sections to other addresses, but within a section instructions and data are
still fixed to their locations. As a result, one can already insert new text-
or data sections. To actually intersperse new instructions within an existing
text section, requires more fine-grained symbolization.

\textbf{Step III: text symbolized assembly.}
Step III takes as input coarsely symbolized assembly and produces assembly where the text sections have been symbolized in a fine-grained fashion (see Figure~\ref{fig:steps_example}).
This step essentially introduces \emph{basic blocks} into the code: blocks of instructions that are always executed sequentially.
The oracle is modeled as:
\[
\mathbf{TextOracle} :: \mathbb{W}_{64} \rightharpoonup \mathsf{FunctionStart} \mid \mathsf{FunctionEnd} \mid \mathsf{BB} 
\]
If $\mathbf{TextOracle}(a)$ is $\mathsf{FunctionStart}$, then address $a$ is a function entry, i.e., the address of the first instruction of some function.
$\mathsf{FunctionEnd}$ means at address $a$ is last instruction of a basic block that ends the function.
This instruction thus can be a return statement, can halt the program, or can call some function that exits the program or does not return.
$\mathsf{BB}$ means that the address is the starting address of some basic block that is not the start or end of a function.
Again, deriving this oracle from a binary is undecidable, in this case because it requires at least resolving of indirections and termination analysis.

Having fine-grained symbolized text sections ensures that 1.) the function
boundary (i.e., entry and exit points) of each function in the binary is known,
and 2.) the CFG of each function is known. Both are crucial in enabling
instrumentability and traversability of the binary. At this stage, one can
intersperse new instructions (e.g., emit a logging message or insert a
null-pointer check).

\textbf{Step IV: stack symbolized assembly.} The stack frames of functions can
be symbolized in fine-grained fashion as well. A stack frame is a contiguous
memory segment used for storing local objects such as variables, arrays and
structs. Up to this step, there is no information available on how the stack
frame is structured, i.e., how the contiguous memory can be broken down into
objects. It is a notoriously hard and undecidable problem to recover structure
of stack frames from a binary~\cite{rose2024modeling}.

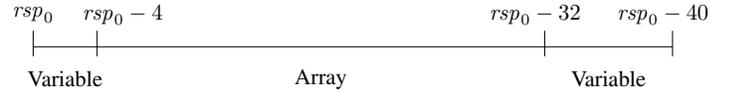
\begin{figure}[h]
\centering
  \scalebox{0.85}{%
  \begin{tikzpicture}%
    \draw [] (0.0,0) -- (10,0);%
    \draw [] (0.0,0.25) -- (0.0,-0.15);%
    \draw [] (1.0,0.25) -- (1.0,-0.15);%
    \draw [] (8.0,0.25) -- (8.0,-0.15);%
    \draw [] (10,0.25) -- (10,-0.15);%
    \node[align=center] at (0.0,0.5)  {$\mathit{rsp}_0$};%
    \node[align=center] at (1.4,0.5) {$\mathit{rsp}_0 - 4$};%
    \node[align=center] at (7.85,0.5)  {$\mathit{rsp}_0 - 32$};%
    \node[align=center] at (9.85,0.5)  {$\mathit{rsp}_0 - 40$};%

    \node[align=center] at (0.5,-0.5)  {Variable};
    \node[align=center] at (4.5,-0.5)  {Array};
    \node[align=center] at (9.0,-0.5)  {Variable};
  \end{tikzpicture}%
}
\caption{Example of stack frame structure.}
\label{fig:stackframe}
\end{figure}

The stack oracle is modeled as:
\[
\mathbf{StackOracle} :: \mathbb{W}_{64} \to \{ \mathbb{W}_{64} \} \\
\]
When some function $\mathtt{f}$ is called with entry address $a$, the stack pointer will have some initial value $\mathit{rsp}_0$.
The stack frame then stores local data at various offsets (see Figure~\ref{fig:stackframe}).
For this example, the oracle will return the set $\{4,32,40\}$.

Having fine-grained symbolized stack frames provides analysis tools with
information necessary to do their work. By inserting the stack frame structure
into the binary, one can, e.g., shuffle the objects on the stack. This is a
form of binary diversification~\cite{bhatkar2003address,bhatkar2005efficient},
a practical technique to prevent bugs from being exploitable. One can also take
a binary with shuffled stack frames and fuzz it, to see if any behavior changes
with respect to the original. If so, that would indicate some out-of-bounds
access, or at the least some violation of how the stackframe was intended to be
structured by the compiler. Such violations are well-known to be the source of
vulnerabilities. Finally, static analysis tools can do bounds checking, but
this requires them to \emph{know} the bounds of local arrays. That is exactly
the information provided by this oracle.

\textbf{Step V: data symbolized assembly.}
In similar fashion to achieving fine-grained stack symbolization, the data sections can be symbolized in fine-grained fashion as well  (see Figure~\ref{fig:steps_example}).
The data oracle provides this information, and is modeled as follows:
\[
\mathbf{DataOracle} :: \mathbb{W}_{64} \rightharpoonup \mathbb{N}^+
\]
It maps addresses to positive integers.
If, e.g., $\mathbf{DataOracle}(a)$ returns $\mathit{80}$, then the 80-byte region at address $a$ is intended by the compiler to be an independent object.
Thus, after fine-grained data symbolization, one should be able to move and shuffle global objects around in memory without changing the intended behavior of the binary, in similar fashion to stack symbolization.

\begin{myverbbox}{\boxBinary}
0x4000: 
55 48 89 E5 48 8D 0D 20 00 00 00 48 
63 14 A9 48 01 CA FF E2 48 31 C0 E9
00 00 00 00 48 C7 C0 2A 00 00 00 C3
F0 FF FF FF FF FF FF FF F8 FF FF FF
FF FF FF FF
\end{myverbbox}

\begin{myverbbox}{\boxUnsymbolized}
0x4000: push   rbp
0x4001: mov    rbp, rsp
0x4004: lea    rcx, [rip + 0x20]
0x400B: movsxd rdx, [rcx + rbp * 4]
0x400F: add    rdx, rcx
0x4012: jmp    rdx
0x4014: xor    rax, rax
0x4017: jmp    0x4023
0x401C: mov    rax, 42
0x4023: ret
0x4024: F0 FF FF FF FF FF FF FF
        F8 FF FF FF FF FF FF FF
\end{myverbbox}

\begin{myverbbox}{\boxCoarseSymbolized}
text+00: push   rbp
text+01: mov    rbp, rsp
text+04: lea    rcx, [data+00]
text+0B: movsxd rdx, [rcx + rbp * 4]
text+0F: add    rdx, rcx
text+12: jmp    rdx
text+14: xor    rax, rax
text+17: jmp    text+23
text+1C: mov    rax, 42
text+23: ret
data+00: .quad (text+14) - (data+00)
data+08: .quad (text+1C) - (data+00) 
\end{myverbbox}
%TODO: check whether this is data+08

\begin{myverbbox}{\boxTextSymbolized}
BB0: push   rbp
     mov    rbp, rsp
     lea    rcx, [D0]
     movsxd rdx, [rcx + rbp * 4]
     add    rdx, rcx
     jmp    rdx
BB1: xor    rax, rax
     jmp    BB3
BB2: mov    rax, 42
BB3: ret
D0: BB1 - D0
D1: BB2 - D0
\end{myverbbox}

%% =============================================================================
%% == IMPLEMENTATION ===========================================================
%% =============================================================================

\section{ELLF Generation}\label{sec:implementation}

%% ========== Introduction

All the information that is needed to generate the oracles is available at some
time during building, some of the information is preserved in the resulting
ELF/DWARF file.
We require the compiler to output some extra information that is not generated
by default.
We modify the compiler arguments to include \texttt{-g} to always generate
DWARF information, \texttt{-fbasic-block-address-map} to emit information about
the basic blocks in the binary, \texttt{--emit\allowbreak -jump\allowbreak
-table\allowbreak -sizes\allowbreak -section} to detect which pointers belong
to the same jump table, and the linker arguments to include
\texttt{-Wl,-Map=<name>.map} for a mapping from the original object files to
their locations inside the binary.
The debugging information is discarded after metadata generation as it is not
needed in the final binary or for disassembling.
Additionally, we add symbols to the start of every section and store
relocations to these symbols to improve the reliability of the linker map.

Table~\ref{tab:implementation-information-sources} contains the sources for
the metadata for each step from Section~\ref{sec:overview}.
Each source is annotated by the stage in which it is available:
Can we get the information in a standard ELF file (\emph{ELF}),
only binaries with debugging information (\emph{DWARF}) or is the information
only available during compiling and linking (\emph{CC/LD})?

%% ========== Disassembling

\textbf{Disassembling.}
For basic disassembling, we need to know the addresses of all instructions
in the binary.
This is information that is not available by default during any stage of
compilation.
Clang supports an option to emit information about the basic blocks in the binary.
Inside a basic block, all bytes must be instructions since there are no jumps to
skip over some bytes.
Conversely, every byte that does not fall within any basic block must be padding
or data.
From this information, we generate the $\mathbf{InstructionOracle}$.
Storing the address of every instruction in the metadata would be very expensive.
Instead, we compress this information into instruction regions that contain the
address of the first instruction and the number of uninterrupted, sequential instruction.
This information is more coarse than the basic block information, multiple basic blocks
can be collapsed into a single instruction region.
Additional metadata that is necessary to recompile the binary, such as the dynamic
libraries required, is available in the standard ELF headers and does not need to
be included in the metadata.
%
%\[
    %\mathbf{InstructionMetadata} :: [
        %\langle \mathit{start} :: \mathbb{W}_{64},
                %\mathit{count} :: \mathbb{W}_{64} \rangle
    %]
%\]

%% ========== Coarse

\textbf{Coarse Symbolization.}
For the coarse symbolization, we need to know which operands in instructions and
which bytes in the \texttt{.data} sections are pointers to other locations, and which
are pure data.
In a position independent binary, resolving which instructions contain pointers
to symbols is trivial as they do not use absolute addresses.
In position dependent binaries, this information is accessible through
\emph{relocations}.
%
%
% For some reason this forces a page break?
{
  %%%%%%%%%%%%%%%%%%%%%%%%%%%%%%%%%%%%%%%%%%%%%%%%%%%%%%%%%%%%%%%%%%%%%%%%%%%%%%
  %% SYMBOLS AND SECTIONS %%%%%%%%%%%%%%%%%%%%%%%%%%%%%%%%%%%%%%%%%%%%%%%%%%%%%%
  %%%%%%%%%%%%%%%%%%%%%%%%%%%%%%%%%%%%%%%%%%%%%%%%%%%%%%%%%%%%%%%%%%%%%%%%%%%%%% 
  
  \newcommand{\SrcFull}{$\checkmark$}
  \newcommand{\SrcFullStripped}{$\checkmark\kern-0.5em\checkmark$}
  \newcommand{\SrcPartial}{$\triangle$}
  \newcommand{\SrcFullWithMod}{\SrcFull}
  \newcommand{\SrcFullInPIE}{\SrcFullStripped$^1$}
  \newcommand{\SrcFullInline}{\SrcFull$^2$}

  \newcommand{\SecLLVM}{\texttt{.llvm\_bb\_addr\_map}}
  \newcommand{\SecJumptab}{\texttt{.llvm\_jump\_table\_sizes}}
  \newcommand{\SecDynamic}{\texttt{.dynamic}, \texttt{.dynstr}}
  \newcommand{\SecCode}{\texttt{.text}}
  \newcommand{\SecDebug}{\texttt{.debug\_info}}
  \newcommand{\SecReloc}{\texttt{.rel.*}, \texttt{.rela.*}}
  \newcommand{\SecSymtab}{\texttt{.symtab}}
  \newcommand{\SecEHFrame}{\texttt{.eh\_frame}}
  \newcommand{\SecPLT}{\texttt{.plt}, \texttt{.dynsym}, \texttt{.dynstr}}

  %%%%%%%%%%%%%%%%%%%%%%%%%%%%%%%%%%%%%%%%%%%%%%%%%%%%%%%%%%%%%%%%%%%%%%%%%%%%%%
  %% ROWS %%%%%%%%%%%%%%%%%%%%%%%%%%%%%%%%%%%%%%%%%%%%%%%%%%%%%%%%%%%%%%%%%%%%%%
  %%%%%%%%%%%%%%%%%%%%%%%%%%%%%%%%%%%%%%%%%%%%%%%%%%%%%%%%%%%%%%%%%%%%%%%%%%%%%% 

  \newcommand{\TableHeaderRow}{
    \textbf{Purpose} & \textbf{Information} & \multicolumn{3}{c|}{\textbf{Present In}}          & \textbf{Location}  \\
                     &                      & ~\textbf{ELF}~ & \textbf{DWARF} & \textbf{CC/LD}  &                    \\
    \hline
  }
  \newcommand{\TableDisassemblingRow}{
    \multirow{3}{*}{\textbf{Disassembling}}
    & Instruction Addresses            &                    &                & \SrcFull{} & \SecLLVM{}    \\
    & \texttt{\_start} Symbol Address  & \SrcFullStripped{} &                &            & ELF Header    \\
    & Required Dynamic Libraries       & \SrcFullStripped{} &                &            & \SecDynamic{} \\
    \hline
  }
  \newcommand{\TableCoarseSymbolizationRow}{
    \multirow{4}{*}{\textbf{Coarse Symbolization}}
    & Pointer-To-Text Operands  & \SrcFullInPIE{}  &               & \SrcFull{} & \SecCode{},  \SecReloc{} \\
    & Pointer-To-Data Operands  & \SrcFullInPIE{}  &               & \SrcFull{} & \SecCode{},  \SecReloc{} \\
    & Pointer-To-Text Variables &                  & \SrcPartial{} & \SrcFull{} & \SecDebug{}, \SecReloc{} \\
    & Pointer-To-Data Variables &                  & \SrcPartial{} & \SrcFull{} & \SecDebug{}, \SecReloc{} \\
    \hline
  }
  \newcommand{\TableTextSymbolizationRow}{
    \multirow{5}{*}{\textbf{Text Symbolization}}
    & Internal Function Addresses               & \SrcFull{}         &               &                   & \SecSymtab{}               \\
    & Statically Linked Function Addresses$^2$  & \SrcFull{}         &               &                   & \SecSymtab{}               \\
    & Dynamically Linked Function Addresses     & \SrcFullStripped{} &               &                   & \SecReloc{}, \SecPLT{}     \\
    & Internal Label Addresses                  & \SrcFullInPIE{}    &               & \SrcFull{}        & \SecCode{}, \SecLLVM{}     \\
    & Jump Table Information                    & \SrcPartial{}      &               & \SrcFullWithMod{} & \SecReloc{}, \SecJumptab{} \\
    \hline
  }
  \newcommand{\TableStackSymbolizationRow}{
    \multirow{2}{*}{\textbf{Stack Symbolization}}
    & Local Variable Locations  & & \SrcFull{} & & \SecDebug{} \\
    & Local Variable Types$^3$  & & \SrcFull{} & & \SecDebug{} \\
    \hline
  }
  \newcommand{\TableDataSymbolizationRow}{
    \multirow{4}{*}{\textbf{Data Symbolization}}
    & Global Variable Addresses             & \SrcPartial{}   & \SrcPartial{} & \SrcFull{} & \SecDebug{}, \SecReloc{} \\
    & Global Constant Addresses             & \SrcPartial{}   & \SrcPartial{} & \SrcFull{} & \SecDebug{}, \SecReloc{}, \SecSymtab{} \\
    & String Addresses                      &                 & \SrcPartial{} & \SrcFull{} & \SecDebug{}, \SecSymtab{}              \\
    & Global Variable Types$^3$             &                 & \SrcFull{}    &            & \SecDebug{}              \\
    \hline
  }
  \newcommand{\TableNamesSymbolizationRow}{
    \multirow{5}{*}{\textbf{Names Symbolization}}
    & Internal Function Names              & \SrcFull{}         &               &            & \SecSymtab{}             \\
    & Dynamically Linked Function Names    & \SrcFullStripped{} &               &            & \SecReloc{}, \SecPLT{}   \\
    & Internal Label Names                 &                    & \SrcPartial{} & \SrcFull{} & \SecLLVM{}, \SecDebug{}              \\
    & Local Variable Names                 &                    & \SrcFull{}    &            & \SecDebug{}              \\
    & Global Variable Names                & \SrcPartial{}      & \SrcPartial{} &            & \SecDebug{}, \SecReloc{} \\
    \hline
  }
  \newcommand{\TableExceptionSymbolizationRow}{
    \multirow{2}{*}{\textbf{Exception Support}}
    & CFI Directives               & \SrcFull{}  & &             & \SecEHFrame{} \\
    & Language Specific Data Areas & \SrcPartial & & \SrcPartial & \SecEHFrame{}, \SecLLVM{} \\
    \hline
  }

  %%%%%%%%%%%%%%%%%%%%%%%%%%%%%%%%%%%%%%%%%%%%%%%%%%%%%%%%%%%%%%%%%%%%%%%%%%%%%%
  %% TABLE %%%%%%%%%%%%%%%%%%%%%%%%%%%%%%%%%%%%%%%%%%%%%%%%%%%%%%%%%%%%%%%%%%%%%
  %%%%%%%%%%%%%%%%%%%%%%%%%%%%%%%%%%%%%%%%%%%%%%%%%%%%%%%%%%%%%%%%%%%%%%%%%%%%%% 

  \begin{table*}[t]
    \centering
    \renewcommand{\arraystretch}{1.2}
    \begin{tabular}{|l|l|c|c|c|l|}
      \hline
      \TableHeaderRow
      \TableDisassemblingRow
      \TableCoarseSymbolizationRow
      \TableTextSymbolizationRow
      \TableStackSymbolizationRow
      \TableDataSymbolizationRow
      \TableExceptionSymbolizationRow
%      \TableNamesSymbolizationRow
    \end{tabular}
    \begin{flushleft}
      \setlength{\leftskip}{20pt}
      \small{
        $^1$: In position dependent binaries, this information is not known. \\
        $^2$: Due to inlining, this information does not always apply. \\
        $^3$: Only the sizes of the types are needed. \\
      }
    \end{flushleft}
    \caption{
      Sources of disassembling and symbolization information in ELF/DWARF binaries.
      The columns ELF, DWARF, CC/LD indicate if the information is collected via the
      final ELF binary, the debugging information in the final ELF binary or
      during the compilation and linking phase.
      Symbols: \SrcFull{} = available,
      \SrcFullStripped{} = available even after \texttt{strip}ing,
      \SrcPartial{} = partially available.}
    \label{tab:implementation-information-sources}
  \end{table*}
}

After compilation, the final addresses of functions and other objects are not known.
Instead, the instructions and variables accessing this information are filled with
zero bytes to be filled in during linking.
Relocations are added as hints for the linker to know which zero bytes need to
be replaced by the addresses of objects.
After linking, the relocations are often no longer needed and discarded.

Clang supports the \texttt{{-}{-}emit-relocs} flags to keep relocations from being
discarded.
However, there have been bugs in \texttt{lld} where relocations got lost even if this
flag is passed.
For this reason, we collect this information directly from the object files,
where the relocations are necessarily present and map their addresses into
the binary using the linker map file.
%
%\[
    %\mathbf{PointerMetadata} :: [
        %\langle \mathit{addr} :: \mathbb{W}_{64},
                %\mathit{addend} :: \mathbb{W}_{64} \rangle
    %]
%\]

If a datum is a pointer according to this metadata then its original type in C/C++
must have been a pointer as well, or it was the target of a function call or
control flow jump.
But not every C/C++ pointer is also in the pointer oracle.
The value of variable like \texttt{volatile int *dst = 0x6000} is a
pointer but it must not be symbolized during the coarse symbolization step.

%% ========== Text

\textbf{Text Symbolization.}
For the text symbolization, we need to add a label to all instructions that are
intended as jump targets.
The only instructions that the compiler intends as jump targets are the first
instructions of basic blocks.
This information is read directly from LLVM's basic block address map.
Additionally, to generate correct assembly code during recompilation, we need
to know which instruction start and end functions.
The start of functions are known via the symbol table in standard ELF but are
more reliably known using the basic block address map.
For the text metadata, we only store the addresses where basic blocks start
and a marker for the basic block that starts a function.
All other information is deduced from there.
%
%\[
    %\mathbf{TextMetadata} :: [
        %\langle \mathit{bb\_start} :: \mathbb{W}_{64},
                %\mathit{is\_entry} :: \mathbb{W}_{8} \rangle
    %]
%\]

%% ========== Data

\textbf{Data Symbolization.}
For the data symbolization, we need to add a label to all global variables.
This information is not available in ELF, but debugging symbols in DWARF
contain debugging information entries for global variables and their types.
Unlike the text symbolization, it is not enough to know the starting address
of global variables to generate valid assembly code for it.
Binaries reuse strings in order to reduce their memory footprint, for example,
if a program uses the strings ``\texttt{Hello World}'' and ``\texttt{World}'',
the binary will only store the former and access the latter via an offset.
Still, DWARF sees these as conceptually distinct variables and contains
incorrect information.
To combat this, we do not keep variables that would overlap with previous ones.
If DWARF misses a variable, our data oracle becomes slightly less precise
but remains sound.
The data that should belong to a missed variable is merged with the variable
directly preceding it.
%
%\[
    %\mathbf{TextMetadata} :: [
        %\langle \mathit{addr} :: \mathbb{W}_{64},
                %\mathit{size} :: \mathbb{W}_{64} \rangle
    %]
%\]

%% ========== Stack

\textbf{Stack Symbolization.}
For the stack symbolization, we need to add a label to all local variables.
This is done in similar fashion to the data symbolization.
The main difference is that local variables belong to a function and are
not available everywhere in the binary.
As such, the metadata for stack symbolization cannot contain purely the
addresses of local variables, it needs to pair them with the address of the
function containing the local variables.
%
%\[
    %\mathbf{StackMetadata} :: [
        %\langle \mathit{func} :: \mathbb{W}_{64},
                %\mathit{off\_from\_rbp} :: \mathbb{W}_{32} \rangle
    %]
    % \]

%% ========== Exceptions

\textbf{Exception Support.}
C++ binaries may define, throw and catch exceptions.
On the binary level, this is implemented via DWARF Call Frame Information (CFI)
entries in the \texttt{.eh\_frame} and the Language Specific Data Area (LSDA) in
the \texttt{.gcc\_except\_table}.
On the assembly level, the CFI entries are represented as directives and
interspersed with normal instructions.
These can be directly reconstructed from the \texttt{.eh\_frame}.
The LSDA needs to be symbolized and references labels from the text metadata for
its landing pads and labels from the data metadata for the virtual tables of the
exception types.
Overall, exceptions do not need any additional metadata that is not already
covered by previous ones.

\section{Evaluation}\label{sec:evaluation}

We evaluate our approach with respect to \emph{correctness} and \emph{overhead}
to validate that the metadata described in Section~\ref{sec:implementation}
enables trustworthy disassembly in practical settings. Our evaluation uses 200
programs from the LLVM test suite, a widely used benchmark for assessing
functional correctness, compilation performance, and binary size. Correctness
is evaluated by executing the test cases. Overhead is quantified in terms of
compilation time, resulting binary size, and execution time of the generated
binaries.
\begin{table*}[t]
  \centering
  \renewcommand{\arraystretch}{1.2}
  \begin{tabular}{l | r r | r r r | r r | c c}
      \textbf{Name} & \multicolumn{2}{c}{\textbf{Compile Time} (s)}          & \multicolumn{3}{c}{\textbf{Binary Size} (KiB)}           & \multicolumn{2}{c}{\textbf{Exec Time} (ms)}  & \multicolumn{2}{c}{\textbf{Test Result}} \\
                  & \texttt{clang} & \texttt{ELLF} & \multicolumn{2}{r}{\texttt{clang}} & \texttt{ELLF} & \texttt{clang} & \texttt{ELLF}          & \texttt{ELLF} & \texttt{ELLF}           \\
                  & \texttt{-O2}  & \texttt{-O2}      & \texttt{-O2} & \texttt{-O2 -g} & \texttt{-O2}      & \texttt{-O2} & \texttt{-O2}             & original & recompiled                   \\\hline
    alacconvert-decode & 17.2 & 26.7 (155.0\%) & 64 & 178 (277.4\%) & 75 (115.8\%) & 18 & 19 (105.0\%) & PASS & PASS \\
    alacconvert-encode & 16.8 & 27.2 (161.9\%) & 64 & 178 (277.4\%) & 75 (115.8\%) & 25 & 24 (99.2\%) & PASS & PASS \\
    burg & 25.8 & 43.6 (168.8\%) & 89 & 218 (244.8\%) & 113 (126.5\%) & 13 & 13 (101.6\%) & PASS & PASS \\
    clamscan & 141.7 & 222.6 (157.1\%) & 977 & 2297 (235.0\%) & 1104 (113.0\%) & 82 & 81 (98.9\%) & PASS & PASS \\
    SIBsim4 & 7.3 & 10.3 (141.3\%) & 69 & 149 (217.5\%) & 74 (107.6\%) & 1001 & 1000 (99.9\%) & PASS & PASS \\
    SPASS & 83.7 & 125.3 (149.7\%) & 787 & 2911 (369.8\%) & 838 (106.5\%) & 3320 & 3377 (101.7\%) & PASS & PASS \\
    aha & 2.2 & 3.8 (173.0\%) & 15 & 24 (162.1\%) & 17 (116.0\%) & 696 & 697 (100.1\%) & PASS & PASS \\
    make\_dparser & 19.0 & 29.0 (152.6\%) & 404 & 665 (164.6\%) & 491 (121.5\%) & 22 & 24 (110.9\%) & PASS & PASS \\
    hbd & 26.9 & 46.0 (170.7\%) & 68 & 248 (365.8\%) & 97 (143.9\%) & 14 & 13 (94.4\%) & PASS & PASS \\
    hexxagon & 9.0 & 13.1 (145.3\%) & 36 & 105 (287.0\%) & 44 (120.6\%) & 3965 & 3820 (96.3\%) & PASS & PASS \\
    \dots & \dots & \dots & \dots & \dots & \dots & \dots & \dots & \dots & \dots \\
    rawcaudio & 2.2 & 3.6 (166.7\%) & 8 & 15 (174.9\%) & 10 (119.8\%) & 10 & 10 (98.0\%) & PASS & PASS \\
    rawdaudio & 2.1 & 3.8 (179.6\%) & 8 & 15 (174.8\%) & 10 (119.8\%) & 7 & 10 (134.7\%) & PASS & PASS \\
    encode & 6.8 & 11.6 (169.2\%) & 19 & 41 (218.6\%) & 23 (122.3\%) & 27 & 28 (104.9\%) & PASS & PASS \\
    toast & 24.9 & 42.9 (172.5\%) & 74 & 187 (252.1\%) & 88 (118.3\%) & 18 & 18 (100.0\%) & PASS & PASS \\
    cjpeg & 72.2 & 117.5 (162.7\%) & 225 & 760 (338.4\%) & 254 (113.1\%) & 12 & 14 (119.3\%) & PASS & PASS \\
    mpeg2decode & 20.4 & 33.2 (162.9\%) & 92 & 182 (199.4\%) & 108 (118.3\%) & 16 & 14 (91.6\%) & PASS & PASS \\
    nbench & 7.8 & 12.3 (156.8\%) & 75 & 165 (221.4\%) & 84 (113.2\%) & 621 & 641 (103.2\%) & PASS & PASS \\
    sim & 2.1 & 2.9 (142.7\%) & 30 & 55 (186.0\%) & 32 (108.5\%) & 1257 & 1257 (100.0\%) & PASS & PASS \\
    frame\_layout & 2.7 & 4.4 (164.5\%) & 145 & 937 (644.6\%) & 193 (132.9\%) & 16 & 17 (106.1\%) & PASS & FAIL \\
    rounding & 1.3 & 2.9 (217.4\%) & 13 & 19 (146.2\%) & 15 (114.6\%) & 8 & 9 (112.3\%) & PASS & PASS \\
    \midrule
    \textbf{Total} & 2875.1 & 4504.6 (156.7\%) & 22534 & 58126 (258.0\%) & 28523 (126.6\%) & 181123 & 178261 (98.4\%) & 198 & 197 \\
  \end{tabular}
    \vspace{0.5cm}
  \caption{
      Comparison between standard \texttt{clang} and our \texttt{ELLF clang}.
      Results are obtained by compiling the LLVM test suite in \texttt{Release (-O2)} mode for both \texttt{clang} and \texttt{ELLF clang}, and \texttt{RelWithDebInfo (-O2 -g)} mode for \texttt{clang}.
      \texttt{Release} mode is taken as the base for the metrics.
      Shown are the compile-time metrics ``Compile Time'' and ``Binary Size'', and the run-time metrics ``Exec Time'' and ``Test Result'' as obtained by the LLVM \texttt{lit} tool.
      }
    \label{tab:evaluation}
\end{table*}

The benchmark set comprises C and C++ programs and reflects realistic
programming patterns, including exception handling and compiler optimizations
that are traditionally challenging for disassembly, such as jump tables.

Our evaluation considers three binary variants: (1) baseline binaries compiled
with \texttt{clang-21}, (2) metadata-augmented binaries produced by
\texttt{ELLF clang}, and (3) binaries obtained by disassembling and recompiling
the metadata-augmented binaries.

\subsection{Experiment}

We use the \textit{MultiSource} benchmarks from version \texttt{22.1.0-rc2} of
the LLVM test
suite.\footnote{\url{https://github.com/llvm/llvm-test-suite/archive/llvmorg-22.1.0-rc2.tar.gz}}
Our evaluation includes all 198 programs that successfully compiled under the
baseline \texttt{clang} configuration and our compilation flags. Two programs
were excluded because they either failed to compile with the baseline compiler
or were incompatible with the selected compilation settings.

While the LLVM test suite provides a standardized and widely used
benchmark for compiler evaluation, it primarily consists of relatively
small, self-contained programs. To assess the behavior of our approach
in more realistic settings, we additionally performed automated
testing on substantially larger applications, including representative
systems such as web servers and image processing tools. For these
programs, we compared the behavior of original and reconstructed
binaries across a range of valid and invalid inputs and did not
observe any deviations. Moreover, the overheads in compilation time
and binary size were consistent with those observed for the LLVM test
suite, suggesting that our results generalize beyond the benchmark
set.

We establish a baseline by compiling all programs in
\texttt{Release} mode using \texttt{clang-21}. In addition, we compile
the programs in \texttt{RelWithDebInfo} mode to quantify the relative
overhead introduced by DWARF debugging information.

To measure the overhead of our approach, we compile the same programs in
\texttt{Release} mode using our \texttt{ELLF clang} wrapper, which invokes the
same \texttt{clang-21} while generating metadata for the resulting artifacts.
This allows us to measure the overhead introduced by metadata generation in
terms of compilation time and binary size.

To ensure that metadata generation does not alter the binaries' behavior, we
execute the full LLVM test suite using the \texttt{lit}
tool\footnote{\url{https://github.com/llvm/llvm-project.git\#subdirectory=llvm/utils/lit}}
on both the baseline and metadata-augmented binaries. This verifies that the
addition of metadata does not alter functional behavior. The reported execution
times additionally allow us to quantify the runtime impact of metadata
integration.

To evaluate the disassembly stage, we process the metadata-augmented binaries
using a fork of the FoxDec~\cite{sound-c-code-decompilation-2020} disassembler
which we modified to consume our metadata. The resulting assembly files are
recompiled using \texttt{gcc-13}. We then re-run the \texttt{lit} test suite on
the recompiled binaries to verify that (1) the disassembly produces valid
assembly code and (2) the reconstructed binaries preserve the original program
behavior.

The metrics collected by \texttt{lit} are summarized in Table~\ref{tab:evaluation}.

\textbf{Functional Correctness.} All 198 original programs pass all test cases.
Similarly, all programs compiled with our toolchain also pass all test cases,
demonstrating that the addition of metadata does not alter functional behavior.
The only exception in the disassembly-and-recompile stage is the program
\texttt{frame\_layout}, which fails a single test case. This failure is
attributable to a rare linker interaction not yet supported by our disassembler
rather than insufficient metadata. These and other limitations are discussed in
detail in Section~\ref{sec:evaluation-limitations}. Overall, these results
indicate that our metadata is sufficient to produce recompilable assembly code
and that the resulting binaries faithfully preserve the behavior of the
original programs.

\textbf{Compilation Time.} Compiling with our toolchain introduces an average
overhead of 57\% relative to baseline compilation. Detailed profiling shows
that approximately 7\% of compilation time is spent on metadata generation
and 5\% on pre- and post-processing of object files; the remainder is spent
on standard \texttt{clang} execution.

\textbf{Binary Size.} Adding ELLF metadata increases binary size by 27\%.
By comparison, including DWARF debug information incurs a larger overhead,
increasing binary size by 158\%.

\textbf{Execution Time.} Execution times of metadata-augmented
binaries are essentially identical to the original binaries. Minor
observed variations fall within expected measurement noise, with
deviations of up to $\pm$10\% relative to the baseline. Notably, we
observe a small average speedup of 1.6\% for metadata-augmented
binaries. This result is based on averaging five independent runs of
the LLVM test suite. Since our approach does not modify the generated
machine code, we attribute this difference to measurement noise and
normal system variability rather than a systematic performance
improvement.

\subsection{Limitations}\label{sec:evaluation-limitations}
\begin{figure}[t]
  \begin{minipage}[t]{0.5\textwidth}
    \textbf{Original Assembly Code}
    \begin{verbatim}
.P1: .quad .L1
.P2: .quad .L2
.L1: .asciiz "Hello World"
.L2: .asciiz "World"
    \end{verbatim}
  \end{minipage}
  \hfill
  \begin{minipage}[t]{0.5\textwidth}
    \textbf{After Merging}
    \begin{verbatim}
.P1: .quad .L1
.P2: .quad .L1+6
.L1: .asciiz "Hello World"
; .L2 got suffix merged
    \end{verbatim}
  \end{minipage}
  \begin{minipage}[t]{0.5\textwidth}
    \textbf{Original Assembly Code}
    \begin{verbatim}
.P3: .L3
.P4: .L4
.L3: .quad some_var
; Assuming the address of some_var
; starts with 0xC0FFEE
.L4: .quad 0xDEADBEEF00C0FFEE
    \end{verbatim}
  \end{minipage}
  \hfill
  \begin{minipage}[t]{0.5\textwidth}
    \textbf{After Merging}
    \begin{verbatim}
.P3: .L3
.P4: .L3+4
.L3: .quad some_var
     .long 0xDEADBEEF

; .L4 got suffix merged 
    \end{verbatim}
  \end{minipage}
  \caption{Example of suffix merging.
    On the top: Suffix merging of string data.
    On the bottom: Suffix merging of pointer data with raw bytes.
  }
  \label{fig:merging-example}
\end{figure}

Our toolchain does not yet fully support some rare binary patterns. In
particular, the disassembler cannot handle merged sections of different types,
and certain unusual switch table layouts are not fully captured by our metadata
generation.

\textbf{Merged Sections.} Linkers may merge multiple read-only data sections
when they contain identical content or when one is a suffix of another, similar
to string merging based on suffixes. Consider the top example in
Figure~\ref{fig:merging-example}: in the original assembly, the strings ``Hello
World'' and ``World'' exist as separate objects (\texttt{.L1} and
\texttt{.L2}). After merging, \texttt{.L2} is removed and referenced as
\texttt{.L1+6}. This does not affect our disassembler, which does not need to
distinguish these objects.

The bottom example in Figure~\ref{fig:merging-example} illustrates a more
challenging case. If a symbolized pointer (the value of \texttt{.L3}) overlaps
with the beginning of the next object (\texttt{.L4}), the linker may merge the
two sections. In this scenario, the value of the merged object depends on the
linker's placement choices, which can differ in the recompiled binary from the
original binary. Such overlaps caused the failure of the \texttt{frame\_layout}
test, and resolving them would require a more sophisticated implementation in
the disassembler.

\textbf{Switch Tables.} High-level switch statements can be lowered by the
compiler in multiple ways. Small or sparse switches may be implemented as
chains of \texttt{cmp} instructions, while larger switches are often lowered
to jump tables. Our toolchain fully supports standard jump tables, extracting
the necessary metadata from the \texttt{.llvm\_jump\_table\_sizes} section.

Switch statements that involve many strings are lowered to lookup tables with
pointers, which are also supported by our metadata generation. However, in rare
cases, compilers generate lookup tables with larger objects for which
relocations do not follow a predictable pattern. In such cases, the metadata
does not provide all information required for correct code generation. In
particular, the disassembler needs to generate pointer values of the form
\texttt{.Lnth\_entry - .Lfirst\_entry}. This requires knowing which entries
belong to the same table, this information is not always known. While this did
not cause any failures in the LLVM test suite, it could potentially affect
other programs with unusual switch layouts.

%% =============================================================================
%% == RELATED WORK =============================================================
%% =============================================================================

\section{Related Work} \label{sec:related-work}

%% ========== Introduction 

Traditionally, disassembling is done by using heuristics to work around the
inherently incomplete information in a binary~\cite{schwarz2002disassembly,andriesse2016depth,pang2021-sok}.
While these approaches are able to achieve impressive results, they can never
be 100\% correct.
We will examine other work that focuses on capturing metadata at build time, and work that aims to rewrite binaries correctly.

%% ========== Ground truth

\textbf{Capturing metadata at build time.}
Alves{-}Foss at al.~\cite{alves2022-inconvenient-truths-of-ground-truth}
note that the existing information in ELF/DWARF is insufficient to be
used as a ground truth for binary disassembling.
They propose recommendations for what a correct ground truth for disassembling
should entail such as the instructions in the \texttt{.text} sections or
the boundaries of functions.
Our metadata is largely compatible with these recommendations.
We found that for disassembling, not all their requirements are necessary,
such as markers for \texttt{noreturn} functions but agree that these
are useful for further analyses.
They say that tracking relocations via compiler modifications is not a
feasible long term solution.
We found that tracking relocations to resolve pointers from the object files
works well in practice, even for position independent code.

Pang et al.~\cite{pang2022-ground-truth-not-easy,pang2021-sok}
modify the \texttt{clang} compiler to emit information such as the
instructions written by the code generation and collect the relocations to
distinguish between symbolized pointers and raw values.
Tracking the compilation from inside the compiler has the advantage that the
result is guaranteed to be correct but it also makes this approach harder to
maintain.
Any changes to \texttt{clang} can break this instrumentation.
Currently, only an older version of LLVM is supported, making it impossible to
generate ground truth for C++20 binaries.

Li et al.~\cite{li2020-on-generation-of-ground-truth}
leverage the listing files emitted by \texttt{gcc} and other compilers
to generate their disassembling ground truth.
Compared to relying on compiler modifications, this makes their approach
applicable to all compilers that are able to emit listing files and is more
resilient towards changes by the compiler maintainers.
As they remark themselves, this also excludes them from using certain compiler
features, such as link time optimizations.

Ince et al. \cite{ince2013-compiler-help} instrument the basic block structure
used during compilation and embed the control flow structure into the binary.
They use this information to aid tools such as
\texttt{Dyninst}~\cite{buck2000-api-runtime-code-patching} as they no longer
need to recover the CFG using static analyses.
Apart from this information (which is comparable to our text metadata), they
do not embed any other metadata.
\\

%% ========== Rewriting

\textbf{Binary Rewriting through Lifting and Recompiling.}
Several approaches aim to provide frameworks that lift a binary to a rewritable IR.
Ramblr lifts binaries into reassemble assembly suitable for patching and instrumentation~\cite{wang2017ramblr}.
Their approach leverages techniques such as control flow graph recovery and content classification (differentiating code from data).
King et al. propose a layout-agnostic binary rewriting
tool called Egalito~\cite{king2020-egalito}.
This tool uses standard disassembling heuristics to parse the binary into
a rewritable format.
As such, these tools are not able to produce correct recompiled binaries 100\% of the
time.

BOLT by Panchenko et al. is a post-link optimizer
leveraging the LLVM framework~\cite{panchenko2019-bolt}.
It extracts a large part of its required information for disassembling directly
from ELF information such as the symbol table and DWARF debugging information.
They mention that relocations can optionally be tracked in the binaries by
passing the \texttt{{-}{-}emit{-}relocs} flag to the linker.
The main differences to our tool are that we aim to minimize trust in information
that could be wrong such as the addresses in the symbol table or the relocations
tracked into the binary by extracting that information directly from the object
files where it must be correct.
Additionally, BOLT works on unstripped binaries with DWARF information and
does the lifting internally while we add a minimal amount of information
into stripped binaries and produce general purpose assembly.

%Superset Disassembly by Bauman et al.~\cite{bauman2018-superset-disassembly}
%uses neither ground truth information from the compilation process nor heuristics
%to solve the undecidable problems inherent to disassembling.
%
%Their tool lifts the binary into a representation of all possible assembly
%programs circumventing the problem altogether.
%
%In contrast to our tool, this will not emit general purpose assembly but
%rather perform the rewrites internally that are consistent with the
%superset representation.

%% =============================================================================
%% == CONCLUSION ===============================================================
%% =============================================================================

\section{Discussion And Conclusion}
\label{sec:conclusion}

\textit{Can ELLF be used to distribute closed-source software?} As explained in
the introduction, we propose here a middle ground binary format that should
make binaries less black-box, but that should also not make it significantly
easier to decompile to source code than a normal stripped binary. We informally
argue that this is indeed the case. Of course, for an ELLF several key problems
that normally are required to be solved when doing decompilation are already
solved. It is thus significantly easier to lift an ELLF to an IR that entails,
e.g., which instructions are to be executed and which not, and which memory
regions are separate objects. That IR, however, is not close in level of
abstraction to source code. Note that in the context of decompilation, not all
source code is created equal. It is, e.g., possible to decompile to ``untyped''
C code (using type punning) with non-descriptive names and with all control
flow implemented by \texttt{goto} statements. It is harder to decompile to
well-structured, humanly understandable and largely architecture independent C
code. With the term ``decompilation'' we here mean decompilation to C source
code that tries to resemble code as it was originally written. Such
decompilation requires several further steps that are \emph{not} made easier:
\begin{description}[leftmargin=0.3cm,labelindent=0cm] \item[Recovering humanly
readable code.] As an example, consider the following x86-64 assembly:
\texttt{mov edx,0x1999999A;} \texttt{imul eax}.
% https://stackoverflow.com/questions/5558492/divide-by-10-using-bit-shifts
The humanly readable equivalent of this snippet is \texttt{edx := eax / 10}
(32-bit integer division by ten). A decompiler can try to recover humanly
readable constructs from the low-level implementations generated by a compiler.
This work is not made easier  by any of the metadata supplied by ELLF.
\item[Recovering syntactic control flow.] ELLF provides control flow
implemented through jumps. To lift this to a representation with syntactic
control flow structures such as if-statements and loops is
non-trivial~\cite{erosa1994taming,brumley2013native,basque2024ahoy} (at least:
not in a way that resembles humanly written code, otherwise it is always
possible~\cite{harel80}). Even though ELLF provides the intended control flow,
how to eliminate all \texttt{goto} statements is not made easier than it is for
regular stripped ELF files. \item[Recovering types.] This is a well-studied but
hard problem~\cite{lee2011tie,noonan2016polymorphic,ziyi23,xie2024resym}. Major
existing binary-level tools such as IDA-PRO and Ghidra try to do type recovery,
but in general they may produce unsound results, or omit types producing holes
in the generated code. Type recovery is not made easier by the metadata added
to the ELLF, other than that it is known how memory segments are structured
into regions.
%\item[Recovering names.] This may refer to names of variables,
%types, or functions. Various research aims at leveraging neural networks or
%LLMs to recover such names from stripped binaries to improve readability and
%human understanding of the lifted
%code~\cite{lacomis2019dire,chen2022augmenting,xie2024resym}. ELLF does not add
%any names anywhere, thereby not making it easier to recover these.
\end{description}

\textit{Is ELLF trustworthy even if DWARF is not?} Our evaluation demonstrates
correctness by showing that the lifted-then-recompiled binary behaves
equivalent to the original. In contrast, correctness of the DWARF debugging
format has been put into
question~\cite{luna2021-who-debugs-debuggers,assaiante2023-where-did-variables-go,li2020-debug-info-optimized-code}.
At first glance, this may seem to contradict ELLF correctness, since some of
the DWARF information is used to generate the ELLF metadata. However, when
DWARF is incorrect it is typically \emph{incomplete}, which has no implication
on correctness of the ELLF metadata. We only rely on DWARF for obtaining
information on variables. If DWARF is incomplete and misses a variable,
fine-grained stack- and data symbolization become a bit coarser but not
incorrect. \\

\textit{Does ELLF work for hand-written assembly?} In general, our
approach does not extend to hand-written assembly. In particular,
identifying instruction boundaries and labels relies on metadata
produced by the compiler pipeline (e.g., LLVM basic block address
maps), which is not available for manually written assembly. More
fundamentally, supporting arbitrary hand-written assembly is not the
goal of this work: ELLF is designed to augment binaries with
compiler-generated information, enabling a setting where disassembly
becomes decidable rather than attempting to solve all challenges of
binary analysis. Nevertheless, limited use cases of hand-written
assembly integrate well with our approach. For instance, embedding raw
data via assembly (as commonly done in systems such as web servers)
does not pose issues. When hand-written assembly code is present, it
does not invalidate the entire binary; rather, only those regions lack
sufficient metadata and may fall back to heuristic disassembly,
without correctness guarantees on that part. Finally, we believe
extending support to hand-written assembly is possible in
principle. Current assemblers do not emit the required metadata, and
our experiments with modifying the LLVM integrated assembler indicate
that such solutions are brittle across compiler versions. In contrast,
\texttt{clang}-emitted metadata has proven relatively stable across
versions in our
experience. \\

\textit{Does ELLF only apply to ELF?} This paper focused on Linux ELF, but all
concepts apply to the Windows PE format and MacOS Mach-O format as well.
Specifically, Section~\ref{sec:overview} essentially describes a specification
of what metadata should be generated at build-time to make the binary liftable
to a handleable IR. Our effort is also largely compiler-agnostic: even though
our implementation is LLVM based, a similar approach could be implemented for
various compilers. \\

%This paper presented a specification of what metadata a binary must contain
%to make it more handleable, and how to generate that metadata at build time.

\textit{Conclusion.} Within the Linux community, there is growing interest in
updating the ELF format. It is one of the topics within The Open Source
Security Foundation (OpenSSF) community of the Linux Foundation, which has
started the Reliable Software Decomposition Special Interest Group (SIG) on
this topic. With this paper, we aim to provide such an effort with a fundament.

%	Linux workgroup for upstreaming

%	FUTURE WORK:
%	New oracle on calling convention:
%	1.) per function call, how many parameters does the function take and which state parts store them. Also, which state part is used to return a value.
%	2.) per function, the size of the stack frame
%	3.) per function call, information on whether the call is expected to return or not
%

%% =============================================================================
%% == APPENDICES ===============================================================
%% =============================================================================

%\clearpage
\bibliographystyle{splncs04}
\bibliography{refs}

\clearpage
\appendix
\section{Artifact Availability}
\label{sec:open-science}

The artifacts underlying this work are available online.

We provide a Docker-based environment to reproduce the evaluation results
described in Section~\ref{sec:evaluation}. The repository includes:

\begin{description}
\item[top-level] Docker environment for reproducing all experiments
\item[foxdec] Fork of the FoxDec disassembler
\item[twelf] Clang-based wrapper for metadata generation
\item[scripts] Scripts to reproduce Table~\ref{tab:evaluation}
\end{description}

The full evaluation results for the LLVM Test Suite are included in the repository.

All materials are available at:
\url{https://osf.io/4ebds/}.
The implementation of the tool is available at:
\url{https://github.com/ssrg-vt/ELLF}
and may contain improvements beyond the version used in this paper.

\end{document}